\documentclass[article]{article}
\usepackage{graphicx}

\begin{document}


\begin{center}
{\bf \large A Systems Model of the Eco-physiological Response of Plants to Environmental Heavy Metal Concentrations}

\bigskip
O.~S. Castillo Baltasar$^a$, 
 N. Dasgupta-Schubert$^a$\footnote{Corresponding author (nita@ifm.umich.mx).
On sabbatical leave at the 
  Dept. of Physics, Southwestern University,
  Georgetown, TX 78628, USA} and C. Schubert$^b$
 
 \begin{itemize}
 \item[$^a$]
{\it 
Institute of Chemical Biology (IIQB)  \\
  Universidad Michoacana de San Nicol\'as de Hidalgo (UMSNH)\\
    CU Ed. A-1, Morelia, Mich. 58060, Mexico\\
nita@ifm.umich.mx
  }
  \item[$^b$]
{\it    Institute of Physics and Mathematics (IFM)\\
   Universidad Michoacana de San Nicol\'as de Hidalgo (UMSNH)\\
   CU Ed. C-3, Morelia, Mich. 58040, Mexico} 
  \end{itemize}
  \end{center}




\noindent
{\bf Abstract:}
The ecophysiological response of plants to environmental heavy metal stress is indicated by the profile of its tissue HM concentrations ($C_p$) versus the concentration of the HM in the substrate ($C_s$). We report a systems biology approach to the modelling of the $C_p$-$C_s$ profile using as loose analogy, the Verhulst model of population dynamics but formulated in the concentration domain. The HM is conceptualized as an ecological organism that `colonizes' the resource zone of the plant cells driven by the potential supplied by the higher HM concentration in the substrate. The infinite occupation by the HM is limited by the eventual saturation of the cellular binding sites. The solution of the differential equation results in the logistic equation, the r-K model. The model is tested for 3 metalophillic plants {\it T.erecta}, {\it S. vulgaris} and {\it E. splendens} growing in different types of substrates, contaminated to varying extents by different copper compounds. The model fitted the experimental $C_p$-$C_s$ profiles well. The $r, K$ parameter values and secondary quantities derived from them, allowed a quantification of the number of Cu binding sites per cell at saturation, the sensitivities (affinities) of these sites for Cu in the 3 experimental systems as well as the extraction of information related to the substrate phyto-availability of the Cu. Thus even though the model operates at the systems level it permits useful insights into underlying processes that ultimately derive from the cumulative molecular processes of HM homeostasis. The chief advantages of the model are its simplicity, fewer arbitrary parameters, and the non-specification of constraints on substrate and plant type.









\vfill\eject\noindent

\section[Introduction]{Introduction}

Metal tolerant plants are species that can thrive in environments contaminated by high concentrations of heavy metals (HM). They constitute an interesting class of plants both from the standpoint of their mechanism of HM homeostasis as well as their potential for use in phytoremediation (Peer et al, 2005). Phytoremediation is an emergent environmental biotechnology that uses plants to clean up the environment (Pilon-Smits, 2005). One of the most important eco-physiological responses of the metal tolerant plant to HM stress, is the profile of its tissue HM concentration ($C_p$) versus the concentration of the HM in the substrate ($C_s$) on which it grows (soil, water or a constructed growth medium). Qualitatively, the nature of the $C_p$-$C_s$ profile has been used to classify HM tolerant plants as metal excluders (ME), metal accumulators (MA), or metal hyper-accumulators (MHA) (Baker, 1981). Amongst this group, the ME plants are possibly the most abundant in the light of the observation that many weeds are ME plants (Wei et al, 2005). Quantitatively, the $C_p$-$C_s$ profile is important for determining the efficacy of the plant for phytoremediation. The phytoremediating efficiency of the plant is given in terms of the time (T) in units of the number of harvest cycles it takes to reduce the initial $C_s$ to a predetermined value, typically taken to be half of the initial $C_s$. The factor that enters into the computation of T is the mass of metal $M_p$ removed by the plant in each harvest which in turn is the product of $C_p$ and the plant biomass $B_p$ (Reeves and Baker, 2000). Conventionally and somewhat simplistically, $C_p$ has been assumed to be a constant irrespective of $C_s$ or to be in a fixed proportion to $C_s$ (Reeves and Baker, 2000; McGrath and Zhao, 2003). However this approach has been shown to result in a serious under-prediction of T (van Nevel et al, 2007). In an attempt to reproduce the $C_p$-$C_s$ response more closely, the recent practice at least for some cases, has been to use regression fitting to obtain a linear relationship between the logarithms of 
$C_p$ and $C_s$ (Zhao et al, 2003; Koopmans et al, 2007; Liang et al, 2009). This approach has served to substantiate the aforesaid under-prediction of T. However, the log scale for both variables reduces the sensitivity of the graph, ``damping'' the profile of an underlying (mathematical) functional dependence emerging at the level of the physiology of the plant organ or whole plant to which the $C_p$-$C_s$ curve pertains. This functional form could yield clues to the implicit mechanism. Furthermore, a purely statistical fit presents difficulties in the assignment of a physical meaning to the regression constants. 

Thus the ideal would be to model the $C_p$-$C_s$ curve mathematically on the basis of the mechanism driving the behaviour. The overall response at the systems level is the cumulative result of the various molecular level processes that determine the metal homeostasis of the plant. These processes include both the abiotic factors such as those responsible for the rhizospheric dissolution of the HM to render it bio-available as well as the biotic factors such as the affinity of the cellular compartments for the HM solute and the protein transporters, chaperones etc. involved in intracellular uptake, trafficking and sequestration of the HM. The so-called reductionist paradigm of modelling identifies then integrates the various molecular level processes to yield the whole plant's response. This is an extremely difficult approach given the complexity of the plant-HM-soil interaction and the fact that much concerning the molecular processes of the phyto-geosphere is not completely known. In an earlier work (Dasgupta-Schubert et al, 2011) we had reported the development of a systems biology inspired approach to the modelling of the $C_p$-$C_s$ profile using as {\it ansatz}, the Verhulst (1838) model of population dynamics. In this model, termed the r-K model, the time domain of the standard population growth model is mirrored by the $C_s$ domain for a fixed period of plant growth: the ``population'' of the HM species within the ``environment'' of the plant i.e. the concentration $C_p$, grows and levels off on account of the interplay between the supply provided by $C_s$ and the resource limitation of the finite number of metal binding sites within the plant. The magnitudes of the intrinsic rate factor `r' and the capacity factor `K' were interpreted in terms of the known physiology of the transport and storage of Cu in ME plants and were shown to be in conformity. This work presents a further investigation of the r-K model that is necessary in order to explore the model's full potential.

We continue to focus on Cu accumulation in plants for two reasons: firstly because Cu occurs amongst the more commonly encountered HM pollutants (Lasat, 2002) and secondly because the homeostasis of Cu in the plant must be tightly regulated, making the response to environmental Cu sensitive. The role of Cu as an essential micronutrient reverts quickly to that of a serious toxin at high environmental concentrations (Yruela 2005). We study the Cu concentration profiles of the roots and aerial parts (the shoots) of the plants {\it Tagetes erecta L.} (marigold) grown in constructed soil, {\it Elsholtzia splendens Nakai ex. F.} grown in soil as well as in hydroponics and {\it Silene vulgaris (Moench) Garcke} grown in soil. The r-K model is made to answer the following questions: (1) Does the model fit this wider data base of Cu ME plants (as compared to the two studied earlier) growing in different substrates with different chemical forms of the Cu contaminant and if so how good are the fits? (2) What can be learned about the nature of Cu uptake by the morphological parts of the roots and shoots in the different cases from the values of the $r$ and $K$ factors (a) for the $C_p$ versus the substrate total Cu concentration ($C_s$) response curve and (b) the $C_p$ versus the substrate solution Cu ($C_s^{\rm sol}$) response curve? (3) What can be inferred about the phytoavailable Cu fraction, implicit in soil total $C_s$, from the $r$ and $K$ parameter values of the $C_p$-$C_s$ and its corresponding $C_p$-$C_s^{\rm sol}$ curves? 
Answers to these questions are essential for a better understanding of the functionality of the r-K model and its applicability to real issues that confront eco-physiologists and phytoremediators such as, the realistic estimate of T as mentioned in the preceding and the question of how labile the HM trapped in the soil is. 

\section[experiment]{Experiment}

\subsection[erecta]{\it T. erecta}

The experiments were conducted in our laboratory whose details are provided in Castillo et al, 2011. The plants were grown in column-mounted pots containing constructed `soil' (2:1 perlite, peat-moss), in a growth chamber. Prior to sowing the seedlings, insoluble copper(II) oxide (CuO) had been dispersed in the soil at the concentrations ($C_s$) of 0 (Control), 500, 1000, 1500, 2000 and 2500 mg of Cu. $kg^{-1}$ dry soil. The average leachate pH for the entire set was 4.95$\pm$0.03. After a 78 day growth period the plants were dried and their roots and shoots separated. The plant parts and leachates were subsequently analyzed for Cu using Atomic Absorption Spectrometry (AAS). 

\subsection[vulgaris]{{\it S. vulgaris} and {\it E. splendens}}

The details of this experiment are presented in Song et al, 2004. The plants {\it S. vulgaris} and {\it E. splendens} were grown for 80 days in a climate-controlled glasshouse. The soils were a total of 30 types of decades old Cu contaminated soils collected from various regions of the world. The chemical forms of the Cu contaminants were as mixed copper oxides, mine tailings copper, and Cu from a $CuSO_4\cdot 5H_2O$ spill. The soil solution had been separately analysed prior to plant growth and its mean pH was 6.7. After harvest the roots and shoots were dried and analysed by Inductively Coupled Plasma Atomic Emission Spectrometry (ICP-AES).

\subsection[hydro]{\it E. splendens}

The details of this experiment are to be found in Yang et al, 2002. The plants were grown hydroponically in a growth chamber for a period of 52 days. The substrate was a modified Hoagland nutrient solution at pH 5.3$\pm$0.2 that had been treated with solutions of $CuSO_4\cdot 5H_2O$ with the Cu concentrations as,
$0.25$, $12.5$, $25$, $50$, $100$, $500$ and $1000\, \mu M$. After harvest the roots and shoots were analyzed for Cu using ICP-AES. 

The chemical forms of Cu in the substrates for all four cases were different. The periods of growth differed but all plants were in their vegetative phase.

\subsection[analysis]{Data analysis}

The substrate Cu concentration unit ($\mu M$) in the experiment 2.3 was converted to mg Cu. $L^{-1}$ to conform to the dissolved Cu concentration, $C_s^{\rm sol}$, unit of the rest of the experiments studied. The published graphs of experiments 2.2 and 2.3 were read off digitally using the image analysis software {\it ImageJ}. 
It needs to be mentioned however that the data in experiment 2.2 showed a wide scatter with several overlapping or occulted data points. This is probably due to the very varied type of the chemical forms of the Cu encountered in the soil with their associated varied mobilizations to the solution (phyto-available) phase. The published graphs of experiment 2.3 had insufficiently graduated y-axes which made the data reduction a difficult operation. The $C_p$-$C_s$ and $C_p$-$C_s^{\rm sol}$data sets were fitted to the generalised r-K and the r-K model equations (eqns. 2 and 4 in section 3.2) using the 
softwares {\it MATHEMATICA} and {\it SigmaPlot} .

\section[r-K]{The r-K Model}

\subsection[basis]{Conceptual basis: molecular processes of plant HM uptake}

Several HM that constitute transition metals of the 4th period of the periodic table are essential mineral micro-nutrients. They are acquired by the root primarily as ions present in the soil solution (the `phyto-available' HM), that interfaces with the root cells (Taiz and Zeiger, 2010a). The concentration of the HM ion in the soil solution, $C_s^{\rm sol}$, is proportional to the concentration of the HM in the immobile soil phase and together they make up the total HM soil concentration, $C_s$. The HM ions permeate the apoplastic vias of the cell wall matrix and reach the external wall of the plasma membrane. The polar groups in the wall and the lipid bi-layer membrane provide adsorptive sites for the ions and polar species (Nobel, 1999). The difference in concentrations and net electric charge of the HM ions across the plasma membrane sets up a membrane electrochemical potential gradient which provides the driving force for cellular filling by the HM (Taiz and Zeiger, 2010b). The HM enters the cell by means of passive and/or active trans-membrane transport processes that are conducted or facilitated by membrane transporter proteins, such as channels and carriers. These protein transporters specify and regulate the trans-membrane (symplastic) absorption of the physiologically important HM (Taiz and Zeiger 2010b) while the physiologically unimportant HM is constrained to remain largely in the neighbourhood of the cell wall and the exterior membrane. 

The higher electrochemical potential consequent to HM build-up in the extra-cellular units causes the increase of intracellular HM. In case of equilibrium between the external and internal compartments, the Nernst equation indicates that the intracellular concentration must increase with the extracellular one through a factor that contains the exponential of the difference between the net charge accumulated in the external and internal compartments (Taiz and Zeiger, 2010b). As the ion from the external membrane face is absorbed into the cytoplasm, the vacant site is filled by another ion from the surrounding soil solution due to the adsorptive quasi-equilibrium between the two phases while the absorbed HM (if in excess) is transported out of the cytosol into the vacuole, making the way for the next intracellular absorption and so on, a process that is supervised by physico-chemical regulation (Taiz and Zeiger 2010b; Hall 2002). For Cu it has been observed (Bernal et al, 2006) that the extracellular accumulation is relatively rapid followed by a slower intracellular accumulation. Experimentally, $C_p$ is obtained as the sum over all cellular compartments for all cells of the given tissue. The aforesaid implies that $C_p$ would rise rapidly the more the $C_p$ initially present (in the extracellular compartments), i.e. $C_p$ is expected to show an initially small then an accelerated progress. Channel proteins that facilitate the inward transport of the HM ions are activated by gates that respond to different external signals such as the change in membrane potential (`voltage-gating') that can occur through changes in the extra-membrane HM concentrations. Once open, they considerably enhance the inward diffusion of the ion (Taiz and Zeiger, 2010b). The sigmoidal response of the ionic current activation of channels to membrane voltage has been observed (Yellen, 1998). Different regulatory mechanisms control this inward flow of ions to different degrees (Plants in Action, 2010; Hall 2002) depending on the nutrient value of the ion so that HM like Cu are relatively more restricted than HM like Fe, especially its symplastic xylem loading onto the photosynthetic organs (Yruela, 2005). At moderate soil solution concentrations, a relatively higher fraction of Cu is found attached to the extra-cellular compartments (Komarek et al, 2010).

\subsection[approach]{Systems approach and model formulation}

We synthesise the array of molecular processes guiding the uptake of the HM by a systems approach to the phenomenon: the solubilized HM species is considered to behave like any ecological organism in that it populates the ``resource'' zone of the plant as constituted by the plant's extra and intra-cellular metal binding sites (MBS).

The potential that drives the HM solute towards the cells of the root is the concentration gradient that exists across the soil solution - plant barrier (Taiz and Zeiger, 2010b; Nobel 1999) which is provided by the magnitude of $C_s^{\rm sol}$. We use the symbol $C_s$ inclusively in this section to mean general soil HM concentration. Once near the cells, the number of HM species that adsorb onto the cells depends on the affinity of the extra-cellular compartments (cell wall and plasma membrane) and the surface area presented for the attachment. The number of HM that subsequently get intra-cellularly absorbed depends on the specific biochemistry of cellular HM absorption of the plant part as described in the section 3.1. That section shows that the number of intra-cellularly absorbed HM depends on the number initially adsorbed in the extra-cellular compartments so that the total number of cellular HM would increase slowly at low $C_s$ representing mostly extra-cellular HM and `accelerate' at higher $C_s$ where the intra-cellular component becomes significant. When all cellular MBS are occupied, saturation will be reached with no further increase of cellular HM with increased $C_s$. The number of such combined MBS and hence the population of the HM for a given $C_s$, would scale with the number of cells or in other words the biomass of the plant organ. The governing factor is the density of occupation which is determined by the concentration or the number of bound HM per unit mass of plant tissue, i.e. the concentration $C_p$. As the plant grows in the presence of $C_s$, more MBS are added but at the same time more HM enters to restore the `equilibrium' concentration. Thus we see that for a given external pressure of $C_s$, the plant acquires a distinctive HM concentration $C_p$ that is typical of the plant's physiology for that particular stage of its growth. This maintenance of the plant tissue HM concentration $C_p$ at a level that is tolerated by the plant, is the systems level manifestation of the various molecular level processes (cf. 3.1) responsible for HM homeostasis. Implicit in this treatment is the assumption that the net number of MBS or the biomass does not change during the process of metal transport into the plant. This is a reasonable assumption because the timescale of biomass growth with respect to solute diffusion over short distances (Nobel 1999) is slow. 

The rate of acquisition of HM concentration $C_p$ by the plant with respect to the environmental HM concentration $C_s$, d$C_p$/d$C_s$, can thus be formulated as driven by 
two mutually antagonistic forces: (1) the unimpeded increase of $C_p$ when the environmental and the metal concentrations within the plant are small and (2) the restriction to further metal ingress due to a substantial concentration of metal already within the plant. Accordingly in the general case,

$$\frac{dy}{dx} = \Bigl(a+b(y-y_0)\Bigr)\Bigl(1-\frac{y}{y_{\rm max}}\Bigr) \hfill \qquad  (1)  $$

Here $x = C_s$, $y = C_p$, so that $dy/dx = dC_p/dC_s$ and the constants $y_0 = C_p^0$ and $y_{\rm max} = K$. 

$C_p^0$ is the innate HM concentration of the plant at $C_s$ = 0 (zero applied dose), in other words the plant HM concentration in normal uncontaminated soils and $K$ is the maximum or saturation value of the HM concentration within the plant. 

When $x\to 0$ so that $y$ is in the neighborhood of $y_0$, 
eqn. 1 reduces to $dy/dx = a$ whose solution is the simple linear equation
$y(x) = ax + y_0$, or, $C_p = a C_s + C_p^0$. It indicates the simple filling of the apoplastic vias by attachment to the extra-cellular compartments where the constant `a' indicates the `rate'. It is similar to the simple linear adsorption isotherm that occurs at low solute concentrations on a solid substrate (Erbil, 2006). While there could be some intracellular absorption, its contribution is too low to be of any significance at these $C_s$ values. 

At intermediate values of $C_s$, the extracellular increase of $C_p$ induces intracellular absorption through the mechanisms as discussed before. The full equation 1 has to be considered wherein the constant `b' relates to the `rate' of the cellular absorption of the HM. The solution with $y(0)=y_0$ is,

$$y(x) = \frac{y_{\rm max}\Bigl({\rm exp}\Bigl[\frac{a+b(y_{\rm max}-y_0)}{y_{\rm max}}\, x\Bigr] -1 \Bigr) + y_0 \Bigl[1+\frac{b}{a}(y_{\rm max}-y_0)\Bigr]}
{{\rm exp}\Bigl[\frac{a+b(y_{\rm max}-y_0)}{y_{\rm max}}\, x\Bigr]  +\frac{b}{a}(y_{\rm max}-y_0)} \qquad (2)
$$

Eqn. 2 shows that when $x$ (i.e. $C_s$) is very large, saturation conditions are reached and $y \to y_{\rm max}$, i.e. $C_p \to K$, the saturation concentration.

The shape of eqn.2 resembles a logistic curve because exponential growth dominates at an intermediate range of $x$.
Since `a' determines the properties of y mainly near the intercept, it is of lesser interest than $b$ and $K$ as far as the HM stressed behavior of plants and phytoremediation are concerned. If eqn. 1 is simplified by neglecting $a$,

$$\frac{dy}{dx}	=	by	\Bigl(1-\frac{y}{y_{\rm max}}\Bigr)   	\qquad					(3) $$

the solution is,

$$y(x)	=	y_0 y_{\rm max} \frac{{\rm exp}(bx)} {y_{\rm max} + y_0\bigl[ {\rm  exp}(bx) - 1\bigr]} 		\qquad		(4) $$

This resembles the familiar logistic equation of the Verhulst model of population growth. Therefore expressing eqn. 4 in the standard form with $x$ and $y$ 
resubstituted by the actual variables, 

$$C_p	=	C_p^0 K \frac{{\rm exp}(rC_s)} { K + C_p^0 \bigl[ {\rm exp}(rC_s) -1\bigr]}		\qquad	(5) $$

where the constant $r$ defines the intrinsic rate of absorption of metal within the plant and $K$ is the carrying capacity. 

Both (2) and (5) fitted the experimental data well (details are presented in the next sections). The fits showed that for the Cu tolerant plant, 
in the given units $a > b$ by slightly less than an order or magnitude. This indicates that the localization of Cu in the apoplasm is significant, corroborating experimental observations (Bernal et al 2006; Komarek et al, 2010). Between eqns 2 and 5, the ratios of the $b$ and $r$ rate factors, the $y_{\rm max}$ and $K$ saturation factors and the $y_0$ and $C_p^0$ innate concentration factors were around 0.98-1.1 for the first two and around 1.5 - 2.0 for the last. This corroborates that 
the constant $a$ in eqn.1 affects principally the intercept while the values of the intrinsic rate and saturation factors are little affected by the simplification to eqn.3. Thus for reasons stated earlier, we continue with eqn. 5, which is analogous in form to the Verhulst model of population growth but in the concentration domain. The analogy with the Verhulst model however does not extend to the underlying mechanism for the behaviour which as we have seen in the foregoing, is driven by a different phenomenon.

$K$ defines the saturated concentration of the HM in the plant tissue and is particularly meaningful for phytoremediation. It is important to remember that it is only the 
$C_s^{\rm sol}$, the phytoavailable fraction, that is relevant to plant HM uptake. $K$ is reached for a certain high $C_s^{\rm sol}$value which corresponds to a certain value of total substrate HM concentration ($C_s$) because the $C_s^{\rm sol}$derives itself from the $C_s$. Thus K will be a constant irrespective of whether the response curve is plotted as $C_p$-$C_s^{\rm sol}$ or $C_p$-$C_s$.

Now the number of HM species that bind to the plant cell, `n', is given by the number of MBS per cell. The total number of the MBS in the tissue is the product of $n$ and the total number of cells. If `m' is the number of cells per unit biomass of tissue, then the total number of cells is the product of $m$ and the plant tissue biomass, $B_p$. The total number of MBS then is the product $nmB_p$. This product in turn is equal to the total number of HM species in the plant tissue, $N$,

$$N =	nmB_p	; \qquad	{\rm or}, \qquad	n =	\frac{N}{mB_p}	=	\frac {1}{m}C_p     	\qquad	(6a) $$

At saturation,	

$$n_K	=	\frac{1}{m} K		\qquad					(6b) $$

Thus K is directly proportional to the number of MBS per cell at saturation, a fundamental quantity regulated by the HM homeostasis mechanism of the cell. The number $n$ will be larger in the cells where absorptive (as against only adsorptive) binding takes place because this means that sites within the volume of the cell (e.g. vacuoles) are also available for metal binding as against only the surface (adsorptive binding). This means that a highly absorbing plant with a high $n_K$ value can achieve a high saturated HM concentration (eqn. 6b). If the biomass value is low however, then the total population of the HM species, $N$, will not be high. This is the situation encountered in several MHA plants (Peer et al, 2005). On the other hand a plant with a moderate value of $n_K$ will have lower $K$ values but if its $B_p$ is high, it will accommodate a substantial population of the HM species. This is the situation encountered in many MA or ME plants (Baker 1981; Pilon-Smits 2005). The magnitude of the $K$ factor will quantitatively determine the suitability of the plant for phytoremediation as anything above that concentration value cannot be attained under the particular conditions. If the conditions change such that $C_s$ becomes too high, the HM homeostatic mechanism within the plant will break down and eqn 5 is not likely to apply at that stage.

	From eqn. 3, at values of $C_p$ much lower than saturation where the value of the rate factor is non-zero and for a plant with a high K factor (the only plants of interest to phytoremediation), the second term in the parentheses may be neglected, resulting in the rate factor,

$$ r	\approx	\frac{1}{C_p}\frac{dC_p}{dC_s}	\approx		\frac{1}{C_p}\frac{\Delta C_p}{\Delta C_s}	\qquad (7) $$

This also follows as an analog of the standard definition of the intrinsic rate factor in the theory of population growth.  
Combining eqn. 7 with eqn. 6a:	 

$$	r	\approx	\frac{\Delta n/n} { \Delta C_s} \qquad	(8)$$

Thus for a unit change in the environmental HM concentration, the rate of increase of the plant HM concentration is equal to the ``per capita'' increase in the number of MBS per cell. This increase is a measure of the propensity of the cellular components that constitute the MBS to become occupied by the HM species. Overall therefore it can be said that the rate factor is associated with the sensitivity of the response of the plant to the level of HM concentrations in the environment. Unlike the case of $K$ however, 
in this case there is a difference between the value of $r$ obtained from the $C_p$-$C_s$ (where $C_s$ now refers to the substrate total HM concentration) and the $C_p$-$C_s^{\rm sol}$ curves (see section 3.3).

	Eqn. 5 defines the r-K model for the HM uptake for plants. It considers the variation of $C_p$ with $C_s$ over a given period of growth for all plants of a given type considered in the set. Different growth periods for the set could result in a family of $C_p$-$C_s$ curves, i.e. time would enter only parametrically. Thus the model is not an explicit kinetic model. We hypothesise that it is possible that for growth periods that do not differ substantially and which concern the same biological stage of growth of the plants of the set (e.g. the vegetative), the response curves could have close values of $r$ and $K$. According to eqns 6b and 8, both the $r$ and $K$ factors depend on the number of MBS per cell ($n$) which is a quantity regulated by the biochemistry of HM homeostasis that is not likely to be very different for plants in the same stage of growth. At a different biological stage of the plant's life, the cellular biochemistry responsible for the HM homeostasis could change resulting in a different magnitude of $n$ which in turn would substantially change the values of $r$ and $K$. The second hypothesis that we make is based on the precept of convergent evolutionary development (McGhee 2011). Plants that utilize the same strategy for HM uptake, e.g. ME plants, are likely to have evolutionarily developed the same type of physiological response vis \`a vis their $C_p$-$C_s$ curves, so that their $r$ and $K$ factors for a given HM are likely to lie within a relatively close range. The two hypotheses are tested in our comparison of the r-K model fits and parameter values for the three Cu ME plants at least two of which have close periods of growth and all are in the vegetative stage. Only the data set for the hydroponically grown E.splendens with a shorter period of growth compared to the others (section 2.3), could probably be an outlier.

\subsection[r]{The r parameter extracted from the $C_p$-$C_s^{\rm sol}$and $C_p$-$C_s$ curves}

Cu and many other HM, are generally found strongly affixed to soil particles (Komarek et al, 2010). However, rhizospheric exudates and ground water are able to produce a certain dissolved fraction of the HM ($C_s^{\rm sol}$) which then becomes suitable for uptake by the roots (phytoavailable). 
This fraction can be conceptualized in terms of the solubility yield factor $\varepsilon$,

$$ \varepsilon	= 	\Delta C_s^{\rm sol} / \Delta C_s 	\qquad							(9) $$

where  $\Delta C_s$ and $\Delta C_s^{\rm sol}$ are the incremental increases of $C_s$ and its corresponding $C_s^{\rm sol}$. 
The factor $\varepsilon$ implicitly includes the phytoavailable fraction of the HM dissolved from $C_s$ (Krishnamurti and Naidu, 2002) where $C_s$ now refers explicitly to the soil total HM concentration. If $r_{C_p-C_s^{\rm sol}}$and $r_{C_p-C_s}$ are the rate factors for the $C_p$-$C_s^{\rm sol}$and the $C_p$-$C_s$ response curves respectively then utilising eqns. 7 and 9, their ratio becomes, 

$$\frac{r_{C_p-C_s^{\rm sol}}}{r_{C_p-C_s}}	=	
\frac{
\frac{1}{C_p}\frac{\Delta C_p}{\Delta C_s^{\rm sol}}
}
{
\frac{1}{C_p}\frac{\Delta C_p}{\Delta C_s}
}
=\frac{1}{\varepsilon}	\qquad (10) $$

The $r_{C_p-C_s^{\rm sol}}$ represents the rate factor driven by the plant only, while $r_{C_p-C_s}$  is the same rate factor modulated by the HM soil solubulization factor 
$\varepsilon$. Infact $\varepsilon$
can be thought of as a scale factor that translates $C_s$ to the $C_s^{\rm sol}$ axis. For sparingly soluble soil metals in the low range of $C_s$ to which the $r$ factor is applicable, the graph of $C_s^{\rm sol}$ against $C_s$ is very nearly linear passing through the origin (Dasgupta-Schubert et al, 2011), in which case 
$\varepsilon$ can be extracted from the slope of the graph. Thus information contained in the $C_p$-$C_s^{\rm sol}$
curve lies embedded in the $C_p$-$C_s$ curve and can be extracted, knowing $\varepsilon$. 
	
For two different plants 1 and 2 growing in two different soils a and b, indicated by the labels (1,a) and (2,b) in eqn. 10, the ratio becomes,
	
$$	
\biggl(
\frac{r_{C_p-C_s^{\rm sol}}(1,a)}{r_{C_p-C_s}(1,a)}
\biggr)
\Big /
\biggl(
\frac{r_{C_p-C_s^{\rm sol}}(2,b)}{r_{C_p-C_s}(2,b)}
\biggr)
= \frac{\varepsilon (b)}{\varepsilon (a)}
\qquad \qquad	(11) $$

Plants mobilise the fixed HM species in soil by exuding dissolving agents such as small organic acids, carboxylates etc. from their roots (Chaney et al, 2010). 
For the same plant growing in the two different soils contaminated by the same HM (plant (1) growing in soil(a) and also in soil(b)), the uptake mechanism would be the same, so that the 
$r_{C_p-C_s^{\rm sol}}(1,a) = r_{C_p-C_s^{\rm sol}}(1,b)$. (The aqueous phase that enters the roots may have different concentrations of the HM depending on the particular rhizospheric conditions of the two soils but the variation of $C_p$ against the $C_s^{\rm sol}$would be the same). The substitution in eqn.11 results in,

$$		\frac{r_{C_p-C_s}(1,b)}{r_{C_p-C_s}(1,a)} = \frac{\varepsilon(b)}{\varepsilon (a)}		\qquad \qquad	(12)   $$

The argument can be extended to two different plants but which belong to the same class of metalophiles (e.g. ME plants), and growing in the two different soils contaminated by the same HM. If convergent evolutionary development operates here, then their intrinsic uptake mechanisms for the same HM would be the same, and if resulting in similar magnitudes of their respective $r_{C_s-C_s^{\rm sol}}$ then, $r_{C_s-C_s^{\rm sol}}(1,a)\approx r_{C_s-C_s^{\rm sol}}(2,b)$. Then from eqn. 11,

$$		\frac{r_{C_p-C_s}(2,b)}{r_{C_p-C_s}(1,a)} \approx \frac{\varepsilon(b)}{\varepsilon (a)}		\qquad \qquad				(13)   $$

Eqns. 12 and 13 imply that the ratios of the $r_{C_p-C_s}$ extracted through the r-K model fits to the $C_p$ versus the soil total HM ($C_s$) curves would yield the information on which soil has the greater dissolved (phytoavailable) content of the HM.  This might have useful applications given the difficulty of directly measuring the phytoavailable content of the HM (Song et al, 2004).
	
The r-K model with its implications as presented in this section, have been tested against the four aforementioned experimental cases and the results are presented in the following section.

\section[results]{Results and Discussion}

\subsection[copper]{The r-K model and copper in plant tissue}

Figures 1-12 and the Table 1 show the results of the fits to the experimental data. In these the $C_s$ stands explicitly for the total Cu concentration in the substrate. The figures for {\it S. vulgaris} have not been included here because they have already been published in Dasgupta-Schubert et al, 2011. However, the values of the model's parameters for {\it S. vulgaris} are re-stated here in Table 1. The $R^2$ values (Figs. 1-12) reveal that the model fits all data rather well. This level of agreement is remarkable given the aleatoric difficulties in reproducing the data from the published graphs (see section 2.4). 
Thus the r-K model as developed in eqn. 5 is a valid quantitative description of the concentration profiles within plants of Cu in the environment and is independent of the manner of expressing the environmental Cu load, i.e. it applies equally to the substrate total Cu as well as to the fraction that is dissolved in it. However, since the process of plant HM uptake concerns only the $C_s^{\rm sol}$, the sections of our subsequent discussion that concern the relevance of the model's parameters to the physiology of Cu uptake by the plant, will focus on what can be learnt from the r-K model fits to the $C_p$-$C_s^{\rm sol}$curves.

We first test the premise of $K$ being independent of the phase ($C_s$ or $C_s^{\rm sol}$) of the HM in the soil (section 3.2). 
Inspecting the entries for {\it T. erecta}, {\it S. vulgaris} and {\it E. splendens} (soil) in Table 1 we see that the K values for the root and the K values for the shoot of a given plant are indeed nearly the same for the $C_p$-$C_s$ and $C_p$-$C_s^{\rm sol}$. Taking the ratio of the $K$ values obtained for each plant organ of each plant for the $C_p$-$C_s$ and $C_p$-$C_s^{\rm sol}$curves, (with the denominator as always the higher value of the pair), the average ratio for the set becomes 0.912. For the variegated experimental data investigated, this level of agreement is quite satisfactory.

The root to shoot translocation factor (TLF) is defined as the ratio of the shoot to the root concentration of the HM (Reeves and Baker, 2000). 
We refine that definition further in the light of the r-K model by expressing it as the ratio of the shoot to root $K$ values obtained in the $C_p$-$C_s^{\rm sol}$curves 
($K_{C_p-C_s^{\rm sol}}$). 
	Instead of a variable concentration $C_p$, $K$ represents the constant, saturated concentration for the plant that is independent of substrate HM concentration, at least till as long as the plant is able to retain its HM homeostasis. From Table 1 we see that for all plants irrespective of the growth medium, the TLF value lies in a narrow range, between approximately 0.1 to 0.2. The fact that the value is less than one indicates that all the plants, regardless of their growth media, are ME plants where the ascent of the HM to the shoot is strongly suppressed and where the root acts as the principal storage unit for the HM (see the K(Root) values in Table 1).
	
The $K$ factors for the root and shoot of {\it S. vulgaris} and {\it E. splendens} (soil) - the two plants that were grown for the same period under identical conditions - closely agree with each other, especially for the shoot. This indicates that at saturation, the number of MBS per cell (eqn. 6) are about the same, indicating a similar mode of Cu sequestration. 
The $r$ factor values however are slightly higher for {\it E. splendens}, especially for the root. It is more convenient to compare between the $r_{C_p-C_s^{\rm sol}}$ (Root) values because the $C_s^{\rm sol}$ gets more modulated by the membrane permeability of the dissolved Cu species in shoot translocation because of the  passage of the Cu solute through more membrane barriers than in root transport (Peer et al, 2005). The $r_{C_p-C_s^{\rm sol}}$ (Root) value for {\it E. splendens} is about twice that for {\it S. vulgaris} that by eqn. 8, indicates that its root MBS likely possess a slightly greater affinity for the Cu solute. 
For the two plants {\it T. erecta} and {\it E. splendens}, the substrates and the specific form of the immobile Cu were different, but their growth periods were very close (sections 2.1 and 2.2). We see from table 1 that their TLF and $r_{C_p-C_s^{\rm sol}}$ (Root) values are nearly the same. 
The latter indicates that the affinities of the root MBS for Cu are nearly the same while the former indicates that their overall 
ecophysiological response is the same. The $K_{C_p-C_s^{\rm sol}}$ (Root) for {\it T. erecta} is about twice as high as for {\it E. splendens} 
which shows that the number of saturated MBS per cell for {\it T. erecta} in the root is higher. On the whole however, these differences (for the root) do not translate to more than a small factor of around 2 which is about the same factor in the maximal difference between the TLF values. It is more the overall similarity of the ecophysiological response, regardless of the type of solid substrate and chemical form of the immobile Cu, as quantitatively articulated by the r-K model, that is noteworthy. This supports the hypotheses outlined in section 3.2 and suggests a convergent evolutionary development in such plants (ME) for the uptake of the HM (Cu).

On the basis of the above it would be supposed that the plant will not suddenly develop a new manner of Cu uptake and storage when presented with a radically different environment. Such an environment is represented by the liquid phase of the substrate and the Cu present as the hydrated ion 
$[Cu(H_2O)_n]^{2+}$ in the case of the {\it E. splendens} grown in the hydroponic medium (table 1). As noted in the preceding, the TLF remains close to the value for the other cases showing that the fundamental physiology of Cu uptake remains the same. Both the root and shoot $K$ indices are however, about an order of magnitude higher than for the case of the same plant grown in soil. Given the immersion of the plant's root in the nutrient solution and the porosity of cell walls, we conjecture that a large part of the accumulated Cu simply lies dispersed in the apoplastic fluid. Upon harvesting and drying, the fluid dries up depositing its contents in the dry plant matter. 
The $r_{C_p-C_s^{\rm sol}}$ (Root) value is lower by a factor of 4.4 which indicates a lower affinity of the cellular MBS for the Cu ion. 

The constant $C_p^0$ obtained from the fits (Table 1) show a large variation. Plant tissue concentrations of copper in uncontaminated soils are generally ~10 mg/kg (Greger, 2004). Of the four plant systems studied in this work, only {\it S. vulgaris} and {\it E. splendens} (soil), were grown in real soil. Hence a meaningful comparison can only be made for those plants. Table 1 shows that for those plants the $C_p^0$ values lie in the environs of 10 mg/kg.

\subsection[copper]{The r-K  model and soil solution copper}

The linearity of the $C_s^{\rm sol}$ vs the $C_s$ plots for the two different solid substrates considered in this work is shown in Figs. 5 and 10 from which the solubility yields 
$\varepsilon$ have been extracted. 
The last two columns of table 1 show the $\varepsilon$ values and the ratios $r_{C_p-C_s}/r_{C_p-C_s^{\rm sol}}$, which according to eqn. 10, must be equivalent. 
We focus on the ratio of the $r$ values for the root because solubulization of the immobile HM into the soil solution is the pre-requisite for root uptake of the HM. Table 1 shows that the values of $\varepsilon$ and the ratio $[r_{C_p-C_s}/r_{C_p-C_s^{\rm sol}}]$ (Root) are equivalent for all plants, thus validating eqn. 10 and the inferences associated with it. 
This equivalence suggests that information on the uptake by the plant root of the phyto-available Cu can be obtained from the root's 
$C_p$-$C_s$ curve and a simple determination of the solubility yield over a limited range of $C_s$ at the mid-lower end of the scale.

For two different plants of the same metalophilic type growing in different soils, eqn. 13 states that the ratios of the rate factors in the $C_p$-$C_s$ curves will be equal to the solubility yield ratios of their soils. To test this, we compared the ratio of the $r_{C_p-C_s}$ factors for the roots of {\it T. erecta} and {\it E. splendens} (soil) with the $\varepsilon$ of their substrates. 
As has been discussed in the preceding, the r-K analysis of these two plants have shown their responses to be quite alike.  
The values at the bottom of table 1 show that indeed the $r$ factor ratios closely match the $\varepsilon$ ratios. 
This validates eqn. 13 and lends support to the inferences associated with it (section 3.3). 
The chief inference that emerges is that metal accumulating plants such as the ME plants discussed here display convergent evolutionary development which makes them respond to HM stress in the environment using a similar set of mechanistic strategies. 
The practical outcome is that if two plants are known to have quantitatively similar responses to the dissolved HM then the ratio of their root rate factors in the 
$C_p$-$C_s$ curves would be in the same proportion as the phyto-available fractions in the two soils. 
The result in Table 1 between {\it T. erecta} and {\it E. splendens} suggests that the Cu in the substrate in which the {\it T. erecta} is growing has the more immobile Cu. 
This is indeed so because the substrate in the {\it T. erecta} experiment (section 2.1) had been spiked with CuO, a refractory oxide of extremely low solubility (Weast and Astle, 1982). 
The CuO might be thought of as approximating the case of extremely aged Cu contaminated soil, where the dissolved Cu species eventually gets mineralised within the soil particles. 
	
The same relationship holds for the same plant growing in two different soils (eqn. 12). In the present work the plant {\it E. splendens} is common to the experiment in soil and in hydroponics. The hydroponics medium is a radically different medium than the soil and as seen earlier, the $r_{C_s-C_s^{\rm sol}}$ (Root) is very different from the rate factor in the soil. Hence eqn 12 cannot be strictly applied in this case. However, we may posit the following imaginary situation: 
	If the hydroponics were to represent a different type of ``soil'', then one would expect the same rhizospheric dissolution processes employed by the plant to take up the Cu. 
Then the $r_{C_p-C_s^{\rm sol}}$ (Root) would be expected to be about the same as in the case of {\it E. splendens} (soil) and the application of eqn. 12 would quantify the $\varepsilon$
of the imaginary soil. The concept of solubility yield does not apply to the hydroponic medium where the Cu is in a completely dissolved state (we may say that the ``solubility yield'' is 1). What this value of $\varepsilon$ would essentially imply then is the fraction of the dissolved Cu that is bio-available. 
Re-designating the $r_{C_p-C_s^{\rm sol}}$ (Root) of the hydroponic medium as the $r_{C_p-C_s}$ (Root) of the imaginary soil (im-soil),
	
$$
r_{C_p-C_s}({\rm Root,\, soil})
\Big /
r_{C_p-C_s}({\rm Root,\, im-soil})
= \varepsilon({\rm soil})/\varepsilon ({\rm im-soil}) 
= 1.3\cdot 10^{-3}/1. (x)
				\qquad (14)
$$
			
Substituting the values of the rate factors from Table 1 we obtain the value of x as 0.24 or the bio-available fraction of the dissolved Cu in the hydroponics medium is about 
$24\%$ of the total dissolved Cu. Even in the idealized case of the HM being present in the completely dissolved form, competing processes such as associations with other species in the solution or absorption by microbes, make the HM less than 
$100\%$ bio-available. Literature reports on the percentage of bio-available copper in bodies of water, such as estuarine water, have determined the percentage to be in the vicinity of $15\%$ (Snyder, 1999).

\section[sumconc]{Summary and conclusion}

The r-K model is a systems level model that attempts to explain the ecophysiological response of the plant to HM stress. While its parameters are connected to the underlying molecular level processes, it does not explicitly investigate the mechanism of HM uptake at the molecular level because that does not fall within its purview. Some semi-reductionist models that attempt to explain the phenomenon at a more microscopic level such as Robinson et al's DSS (Decision Support System) (Robinson et al, 2006) or the Biotic-Ligand Model (BLM) (Paquin et al, 2002) have met with some success but these have been confined to specific cases. A discussion of the systems and reductionist approaches vis \`a vis these models have been done by Dasgupta-Schubert et al. (2011). The significant advantages of the r-K model are the simplicity of its approach, its fewer arbitrary parameters, and the lack of the constraint to particular cases. As shown in the preceding, it applies to all the Cu accumulating plants investigated in this work that were grown in widely differing substrates and for the different morphological parts. We speculate that for HM nutrients with similar physiological roles as Cu, the same applicability might be obtained. Another advantage of the model is its genetic link to the very well-developed model of population dynamics. It holds the possibility of reaching out to its conceptual progenitor to borrow from the latter's richly developed formulation and power, to solve `analogous' problems in the concentration domain. 

Obviously, more experimental data need to be made available to completely test the r-K model. A particularly nagging problem is the dearth of sufficiently dense data in the higher $C_s$ range. Since this range determines the K value it is particularly important to obtain sufficient data points here. Our re-definition of the TLF might also serve as an impetus for this. Future work may envisage looking into the limits of the applicability of the model. We surmise that when the HM regulatory mechanism breaks down, one scenario could be that $C_p$ will not saturate at K but rather display a rapid rise with $C_s$, corresponding to the HM induced damage to the cell's plasma membrane and the consequent cellular ``flooding'' by the HM. For non-nutrient HM for which plants have not developed specific trans-membrane transporters we expect that the intra-cellular absorption will be highly restricted - intrinsic rate values may be very small or negligible and the model may not be followed. Since flow within the apoplasm cannot be restricted, the plant might instead attach the HM to the wall matrix and the $C_p$-$C_s$ profile may perhaps show adsorption-like characteristics.

To conclude, this work has shown that the r-K model fits the experimental data of the three plants, {\it T. erecta}, {\it S. vulgaris}, {\it E. splendens} in their respective growth media, rather well and that useful information regarding the number of metal binding sites per cell, their sensitivity to the affixation of Cu, and the phytoavailability of the Cu in the soil was inferred from the $r$ and $K$ parameter values. The magnitudes of the derived quantities such as the TLF and the $r$ factor ratios, pointed to a similar strategy of Cu uptake in all the systems investigated which supports the conjecture of convergent evolutionary development of such metalophillic plants. 

\section*{Acknowledgements}	
OSCB and NDS are grateful to CONACyT of Mexico for funding their doctoral fellowship (No.169682 ) and sabbatical fellowship in the USA (proyecto no. 173076) respectively.

\vfill\eject

\section*{Figures}

\begin{figure}[h]
\begin{center}
\includegraphics[width=1.15\textwidth]{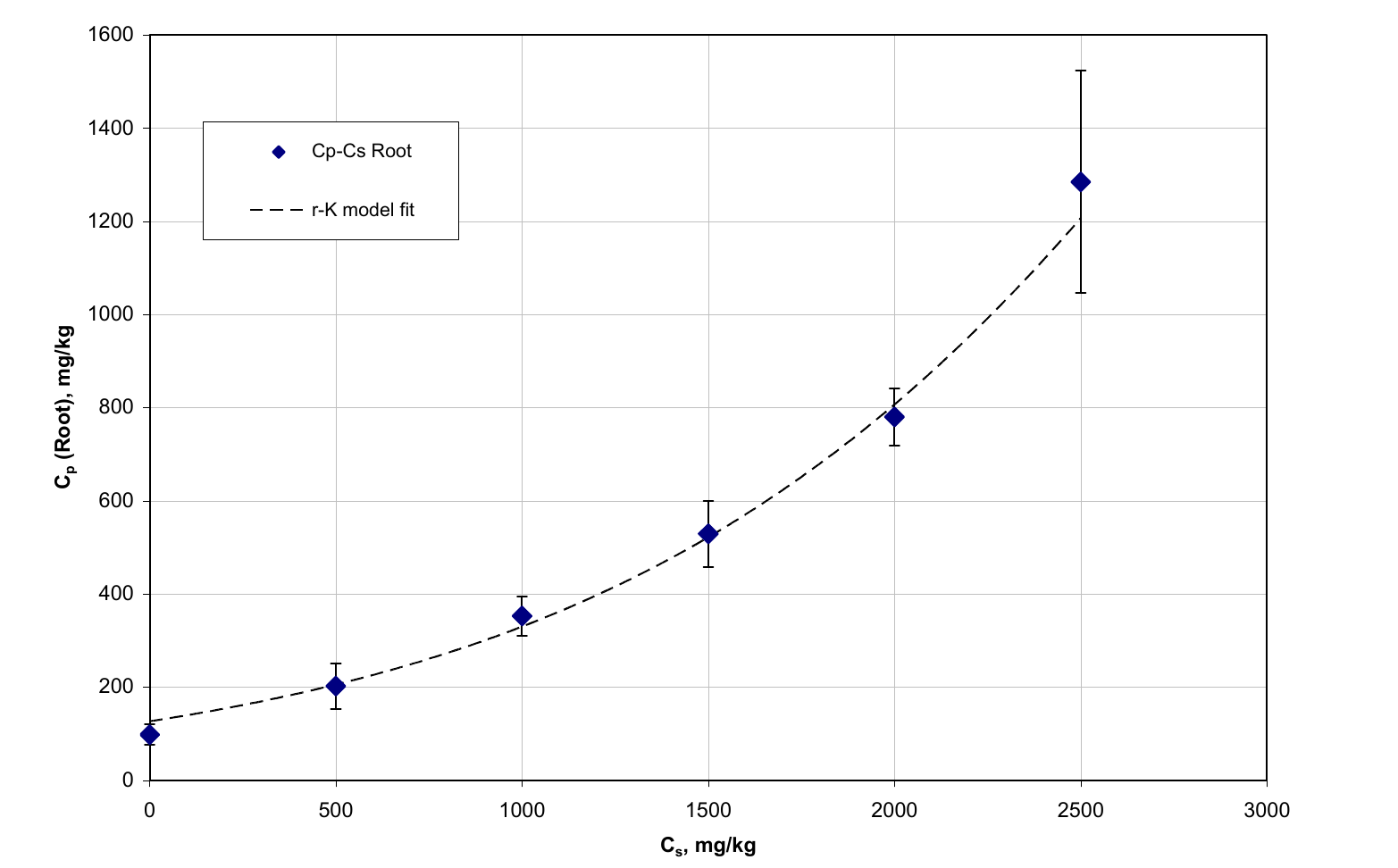}
\end{center}
\caption{The experimental data of the concentration of Cu in the root of the plant {\it T. erecta}, $C_p$ (mg Cu. $kg^{-1}$ dry tissue) for the different doses of Cu in the substrate with the concentrations, $C_s$ (mg Cu. $kg^{-1}$ dry substrate). The Cu dose had been applied in the form of the insoluble CuO. The dashed line is the r-K model (eqn. 5) fit to the data, ($R^2$ = 0.996).}
\label{fig1}

\end{figure}

\begin{figure}[h]
\begin{center}
\includegraphics[width=1.15\textwidth]{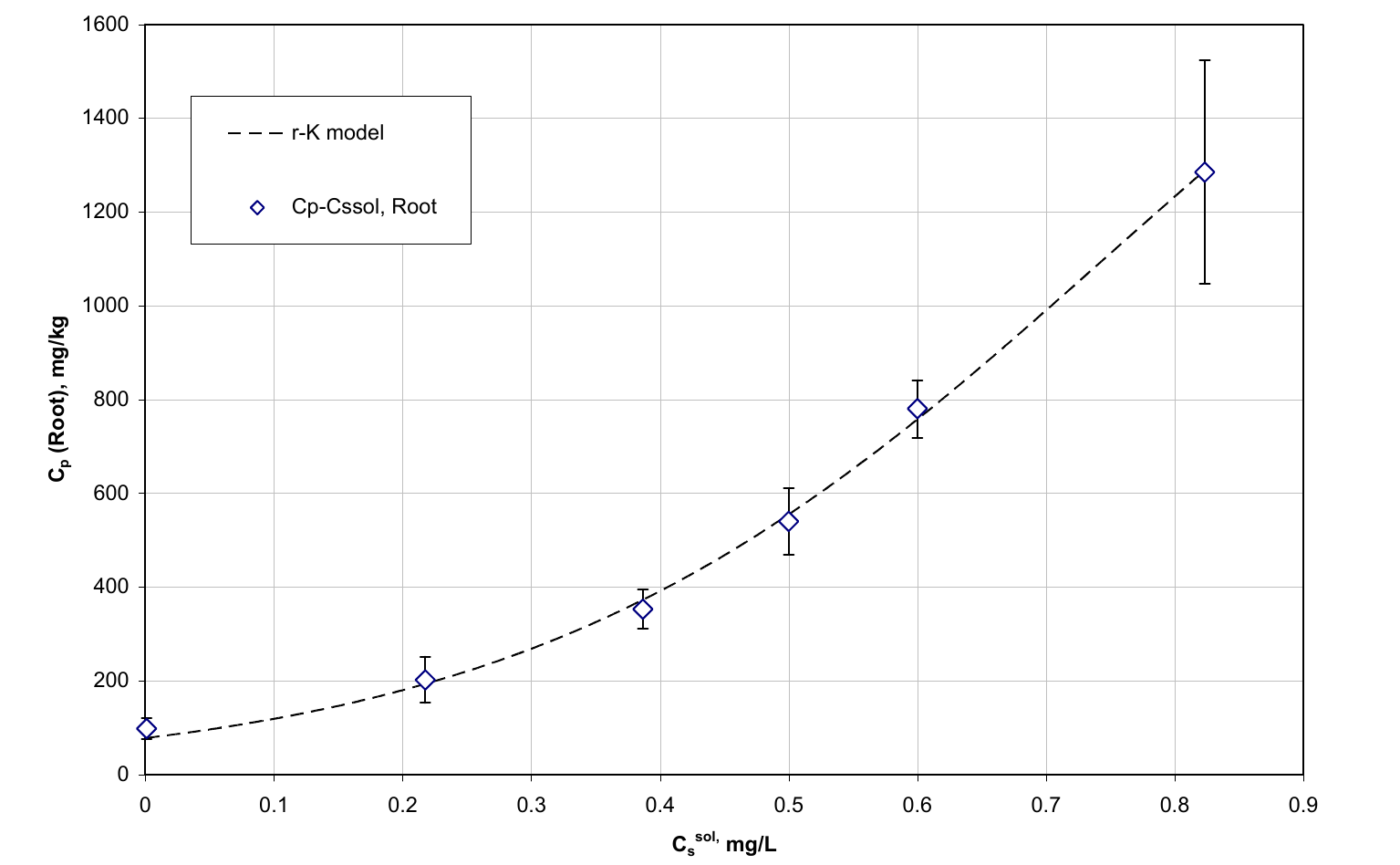}
\end{center}
\caption{The same as fig. 1 except that the substrate Cu concentration (X-axis) now is presented as the substrate leachate (soil solution) concentration $C_s^{\rm sol}$(mg Cu. $L^{-1}$ solution). ($R^2$  = 0.993).
}
\label{fig2}

\end{figure}
\begin{figure}[h]
\begin{center}
\includegraphics[width=1.15\textwidth]{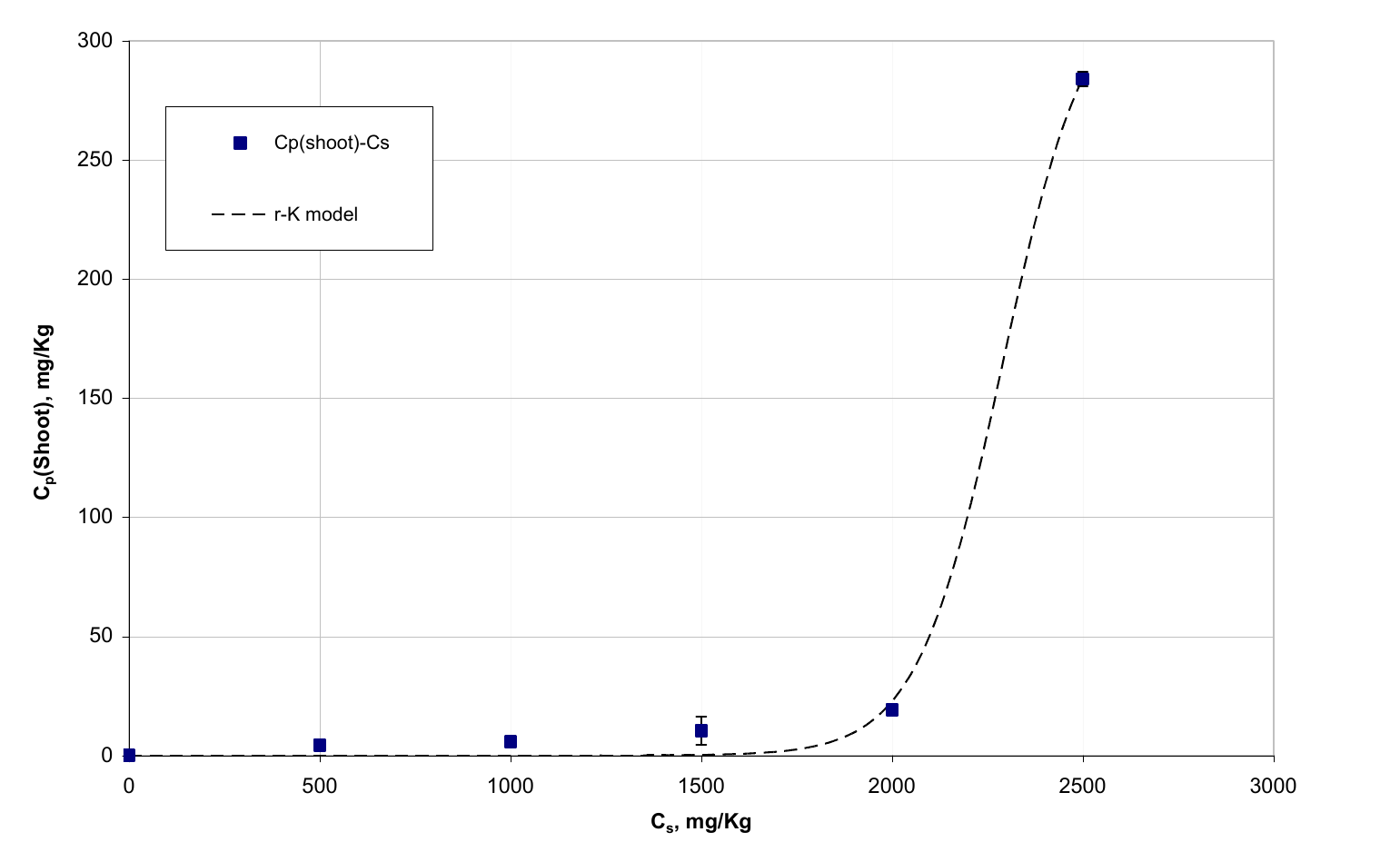}
\end{center}
\caption{The variation of the Cu concentrations, $C_p$, (Y-axis) in the shoot of {\it T. erecta} versus $C_s$ ($R^2$ = 0.995).
}
\label{fig3}

\end{figure}
\begin{figure}[h]
\begin{center}
\includegraphics[width=1.15\textwidth]{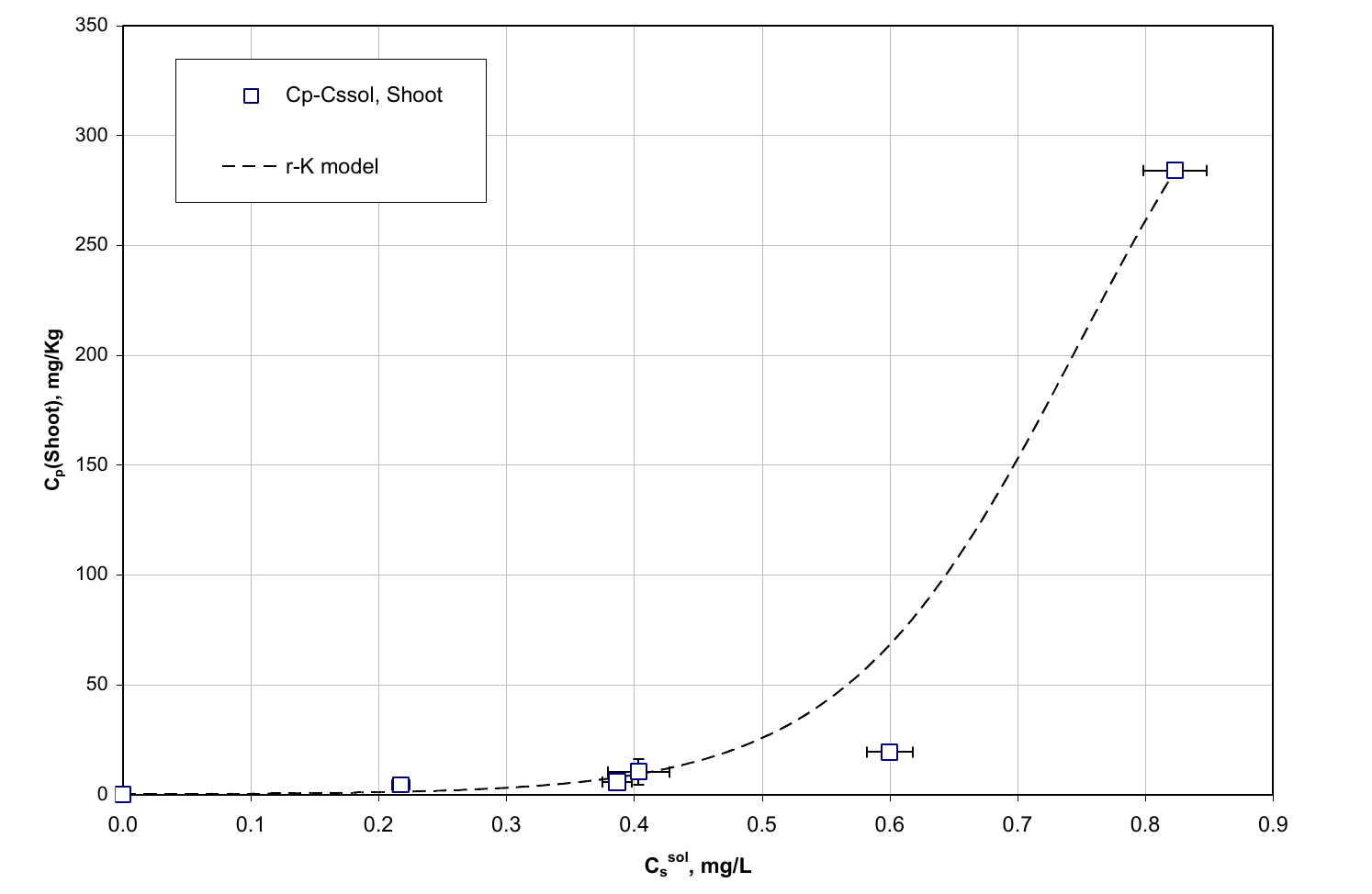}
\end{center}
\caption{The same as fig. 3 but for Cu concentrations in the X-axis expressed as $C_s^{\rm sol}$ ($R^2$ = 0.861).}
\label{fig4}

\end{figure}
\begin{figure}[h]
\begin{center}
\includegraphics[width=1.15\textwidth]{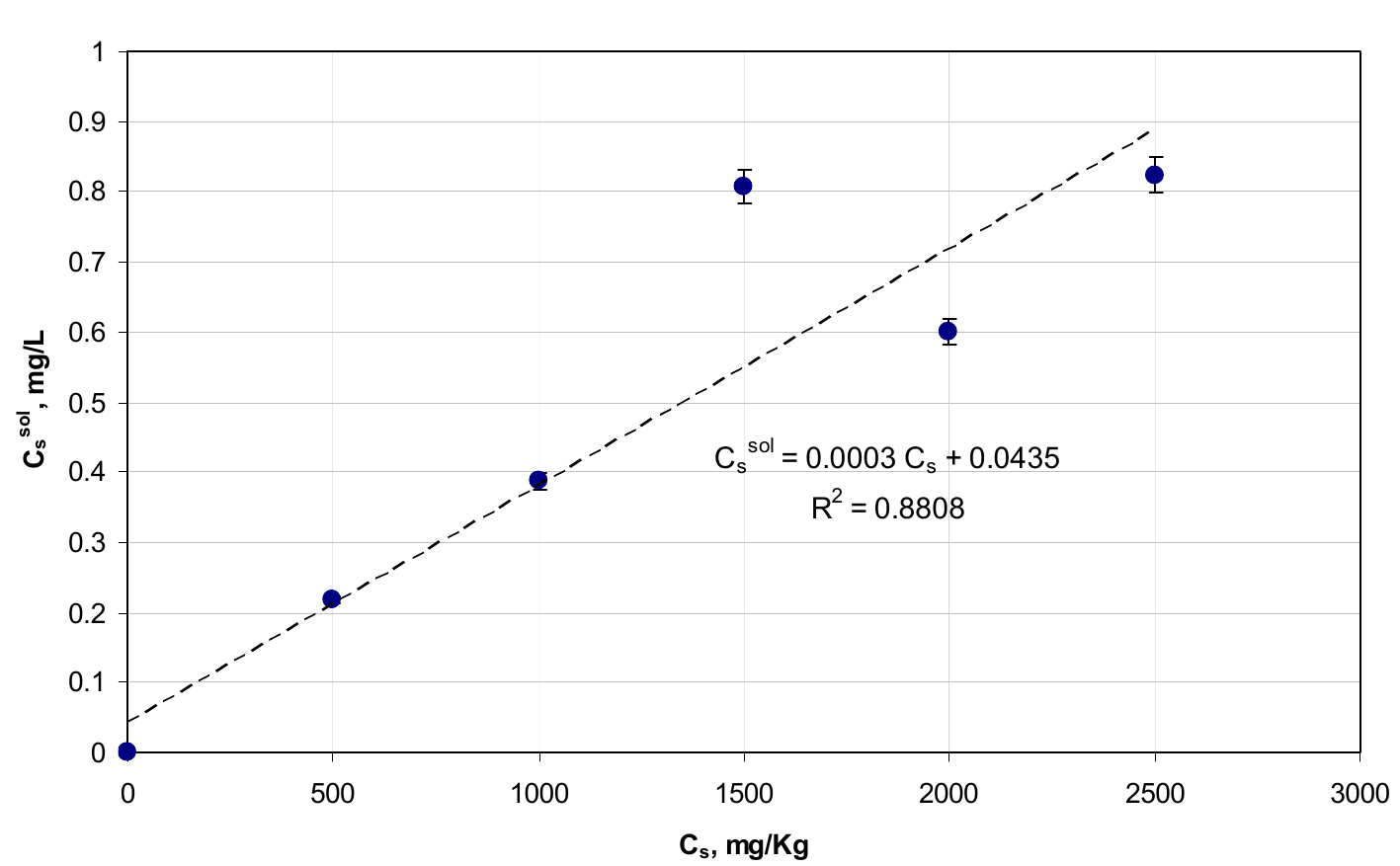}
\end{center}
\caption{The yield of $C_s^{\rm sol}$ with respect to the $C_s$ of the substrate on which the {\it T. erecta} was grown. The solubility yield, $\varepsilon$, is the slope of the fitted straight line, 0.0003.}
\label{fig5}

\end{figure}
\begin{figure}[h]
\begin{center}
\includegraphics[width=1.15\textwidth]{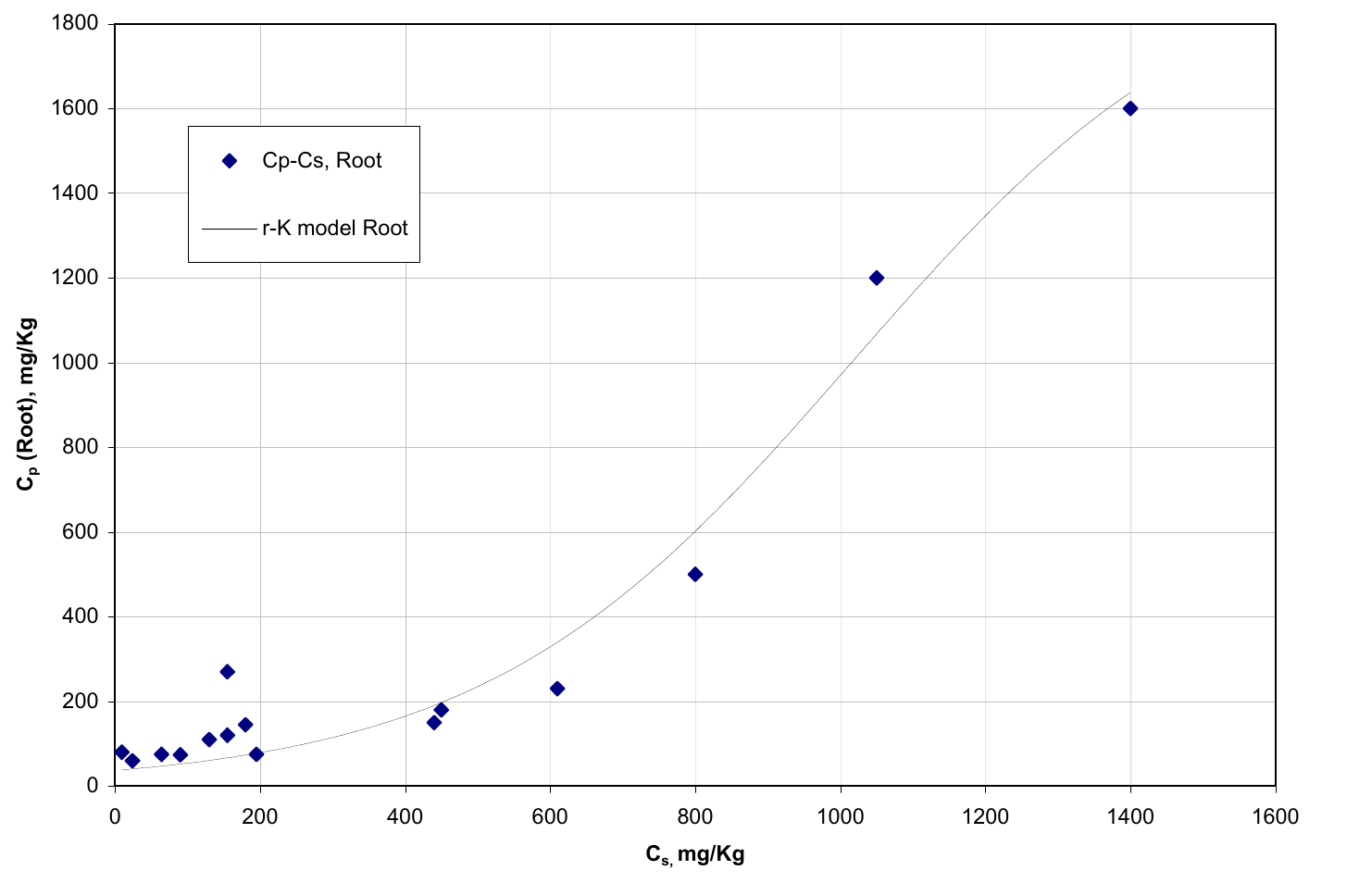}
\end{center}
\caption{The experimental data of the values of $C_p$ in the root of the plant {\it E. splendens}, for the different $C_s$ of the substrate. In this experiment (Song et al, 2004), the substrate was soil with aged contaminations of Cu that arose from various sources (see section 2.2). The dashed line is the r-K model (eqn. 5) fit to the data, ($R^2$ = 0.966). N.B. Song et al's highest $C_p$-$C_s$ data point, with its very high value was a distinct outlier and could not be included in the fit. }
\label{fig6}

\end{figure}
\begin{figure}[h]
\begin{center}
\includegraphics[width=1.15\textwidth]{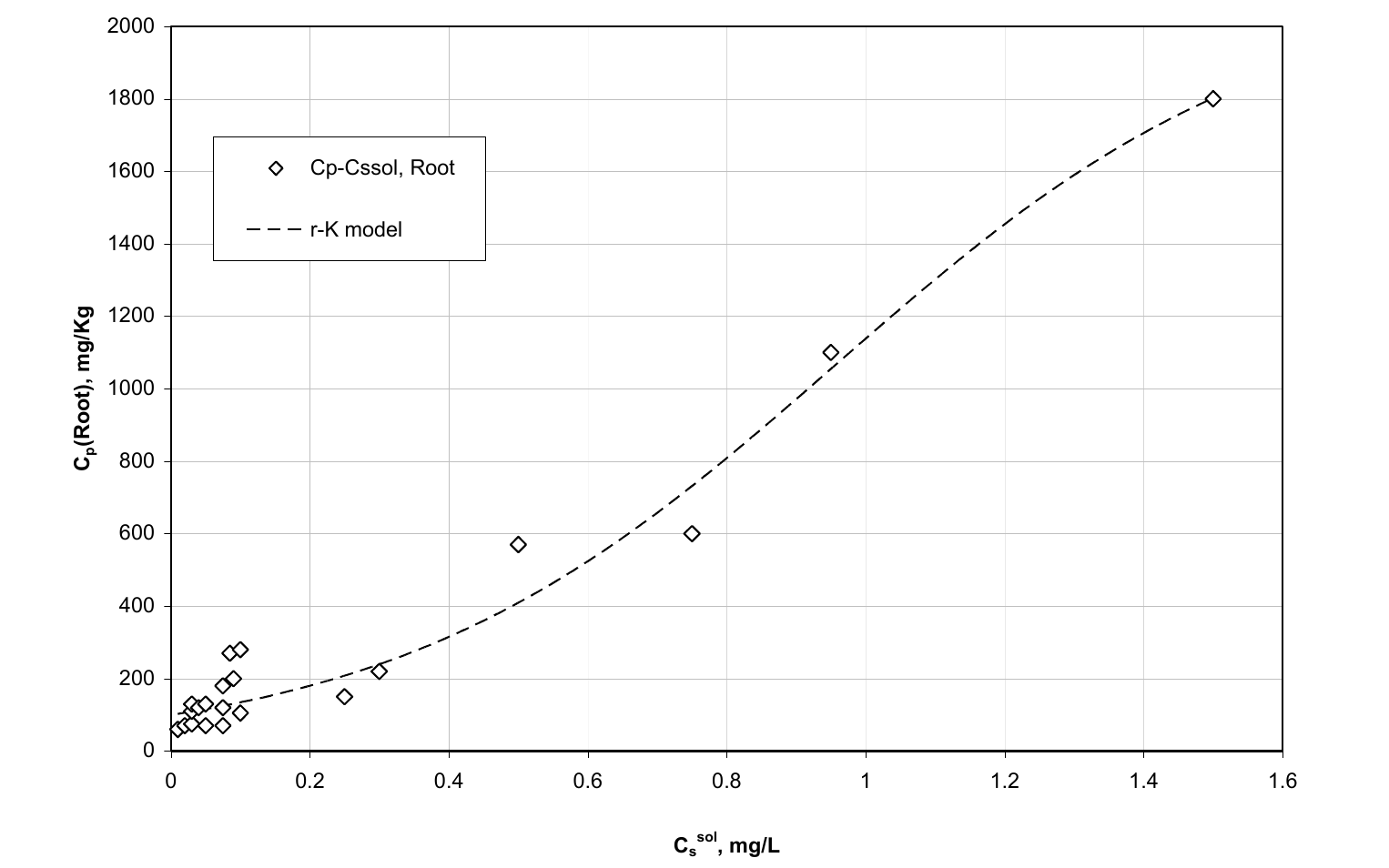}
\end{center}
\caption{The same as fig. 6 but with the X-axis expressed as $C_s^{\rm sol}$, the concentration of Cu in the soil solution ($R^2$ = 0.964). The same caution as in fig. 6 applied to the highest $C_p$-$C_s^{\rm sol}$ data point.
}
\label{fig7}

\end{figure}
\begin{figure}[h]
\begin{center}
\includegraphics[width=1.15\textwidth]{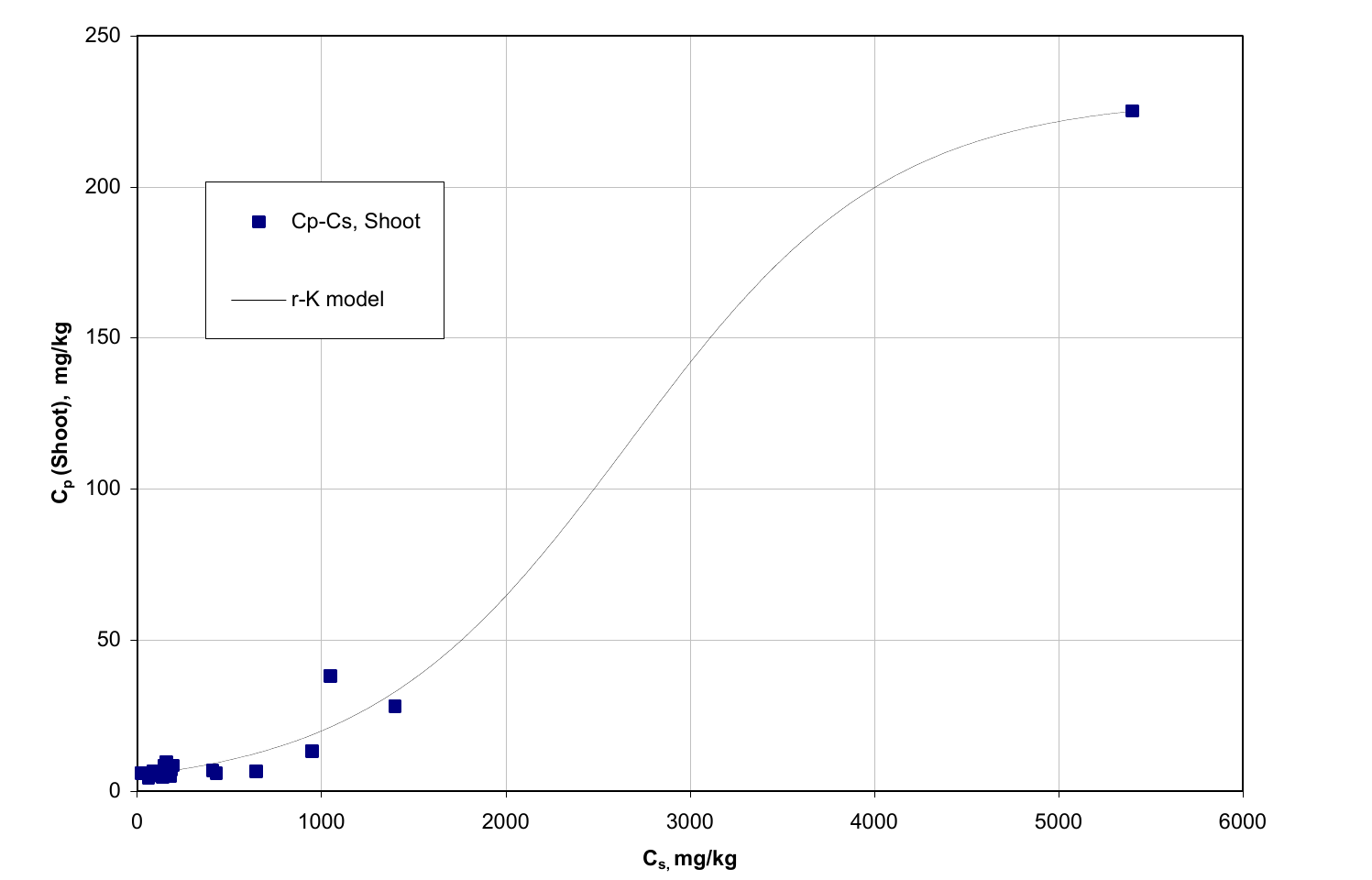}
\end{center}
\caption{The variation of the $C_p$ in the shoot of {\it E. splendens} with respect to $C_s$ ($R^2$ = 0.991).}
\label{fig8}

\end{figure}
\begin{figure}[h]
\begin{center}
\includegraphics[width=1.15\textwidth]{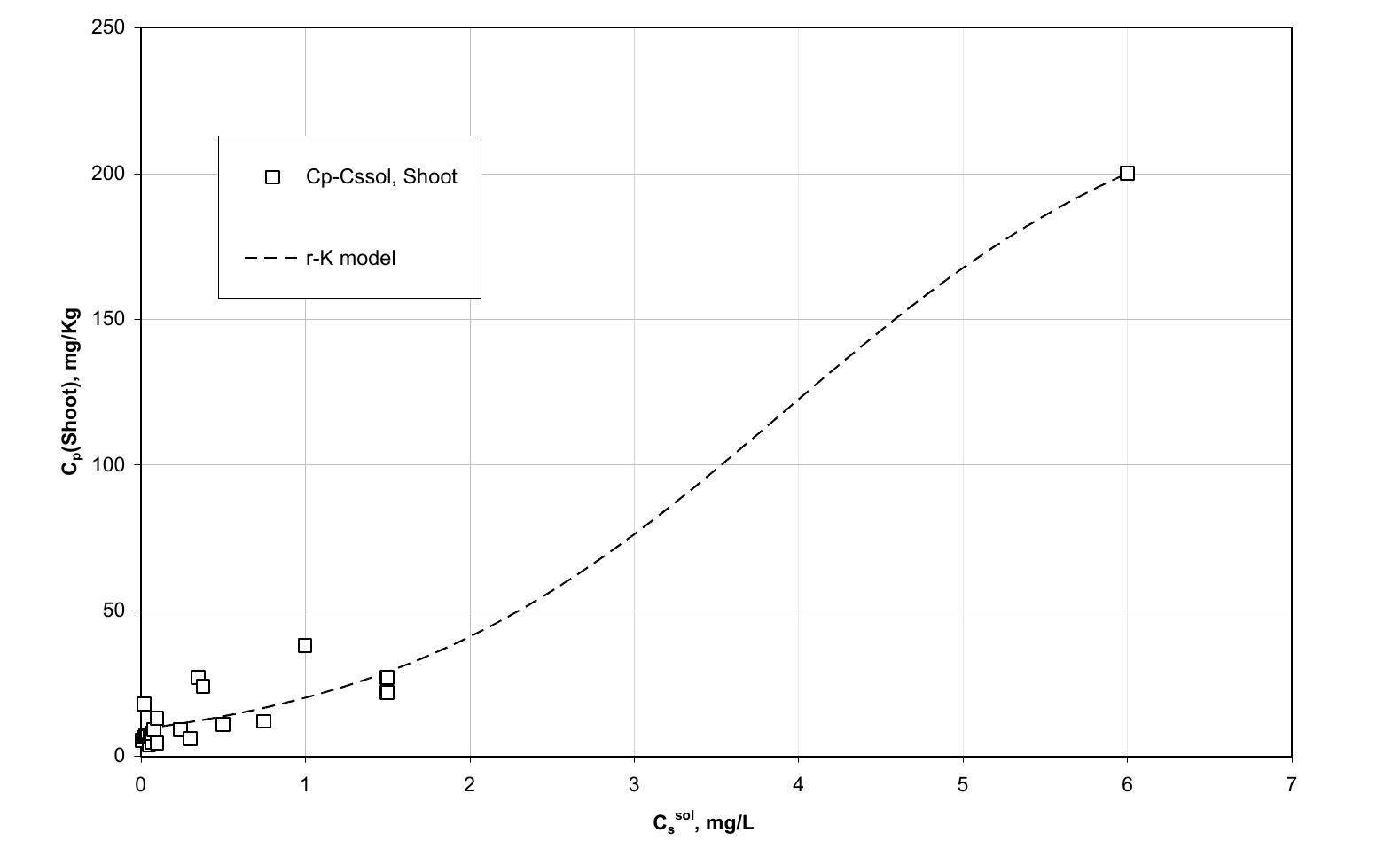}
\end{center}
\caption{The same as fig. 8 but with the X-axis given as $C_s^{\rm sol}$ ($R^2$ = 0.971).}
\label{fig9}

\end{figure}
\begin{figure}[h]
\begin{center}
\includegraphics[width=1.15\textwidth]{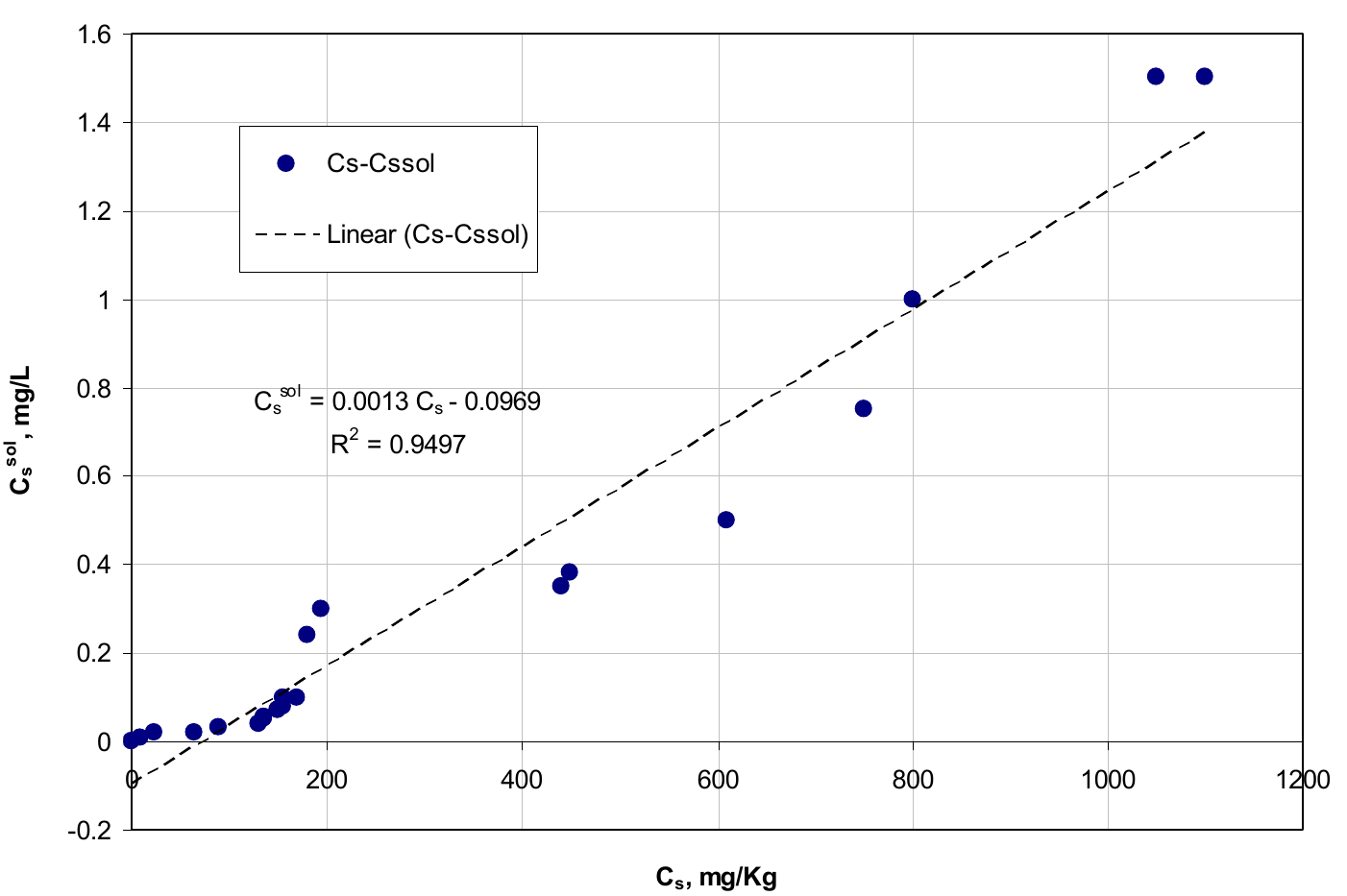}
\end{center}
\caption{The solubility yield data for the soil on which {\it E. splendens} and {\it S. vulgaris} were grown (Song et al, 2004). The dashed line is the linear fit from which $\varepsilon$ 
emerges as 0.0013.}
\label{fig10}

\end{figure}
\begin{figure}[h]
\begin{center}
\includegraphics[width=1.15\textwidth]{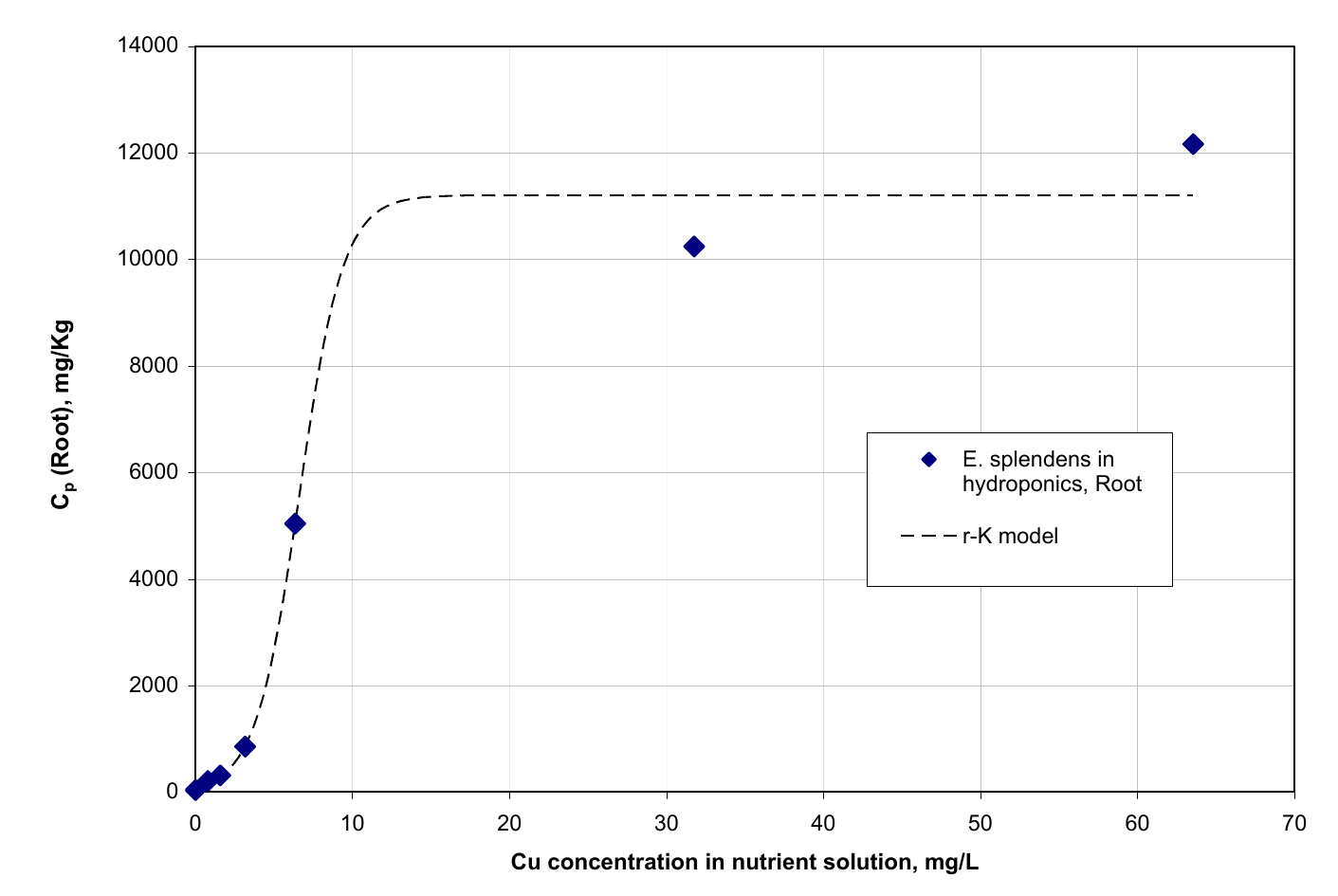}
\end{center}
\caption{The variation of the Cu concentration in the root of {\it E. splendens}, $C_p$, grown hydroponically in nutrient solution (Yang et al, 2002) with varying concentrations of Cu, $C_s^{\rm sol}$ ($R^2$ = 0.988).}
\label{fig11}

\end{figure}
\begin{figure}[h]
\begin{center}
\includegraphics[width=1.15\textwidth]{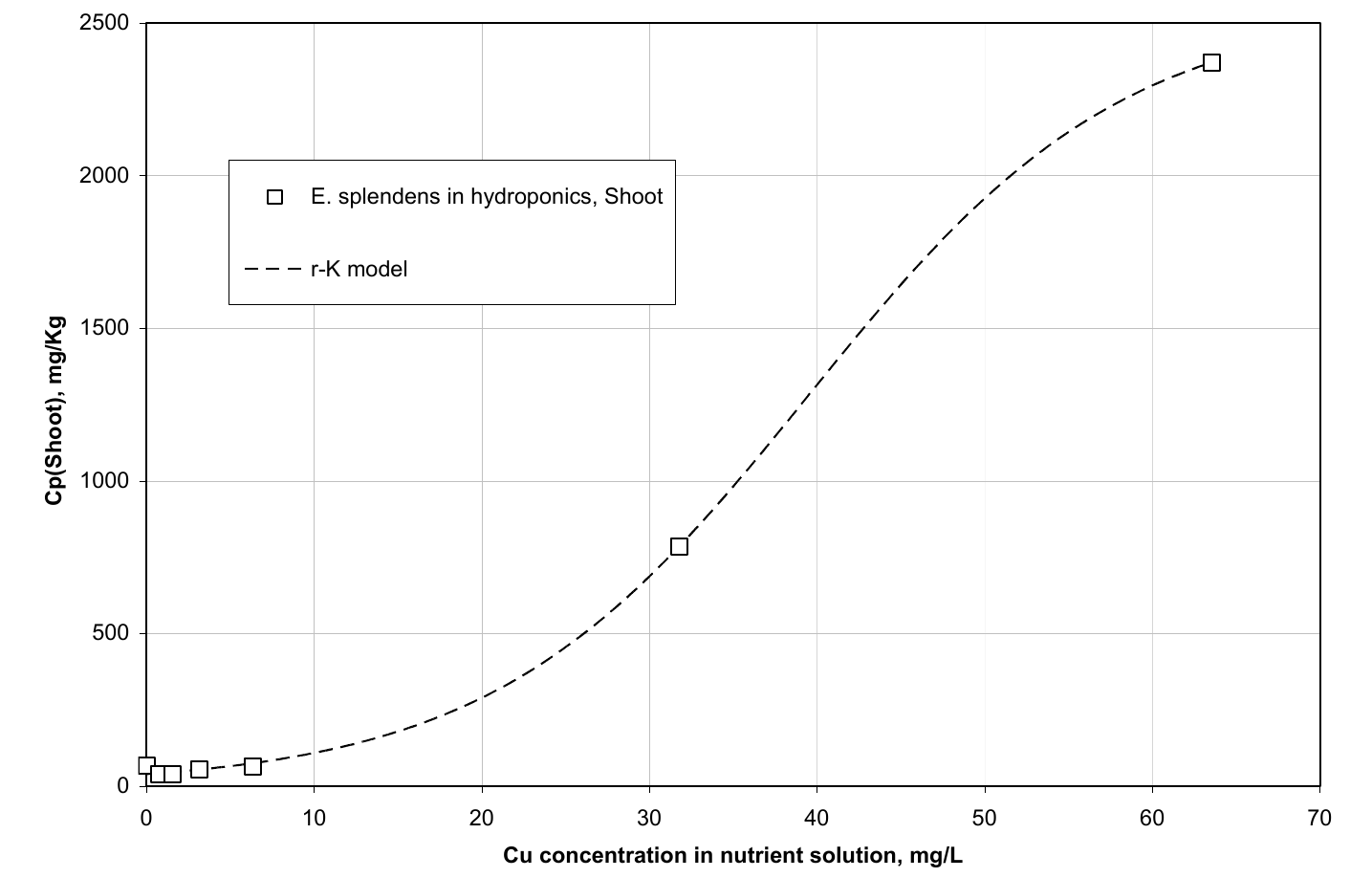}
\end{center}
\caption{The variation of the $C_p$ in the shoot of the hydroponically grown {\it E. splendens} (Yang et al, 2002) with respect to $C_s^{\rm sol}$ ($R^2$ = 0.999).}
\label{fig12}

\bigskip
\bigskip

\noindent
{\Large\bf Table}

\bigskip

Table 1.	The values of the r-K model parameters, $r$, $K$ and $C_p^0$, for the profiles of the copper (Cu) concentrations in plant tissue ($C_p$) versus the soil total Cu concentration ($C_s$), $C_p$-$C_s$, and versus the soil solution Cu concentration ($C_s^{\rm sol}$), $C_p$-$C_s^{\rm sol}$, as obtained for the roots and shoots of the plants {\it T. erecta}, {\it S. vulgaris} and {\it E. splendens} grown in the substrates of constructed soil (Constr. Soil), natural soil and in hydroponics. $\varepsilon$ is the solubility yield of the substrate.

\end{figure}

\begin{figure}[h]

\begin{center}
\includegraphics[width=1.15\textwidth]{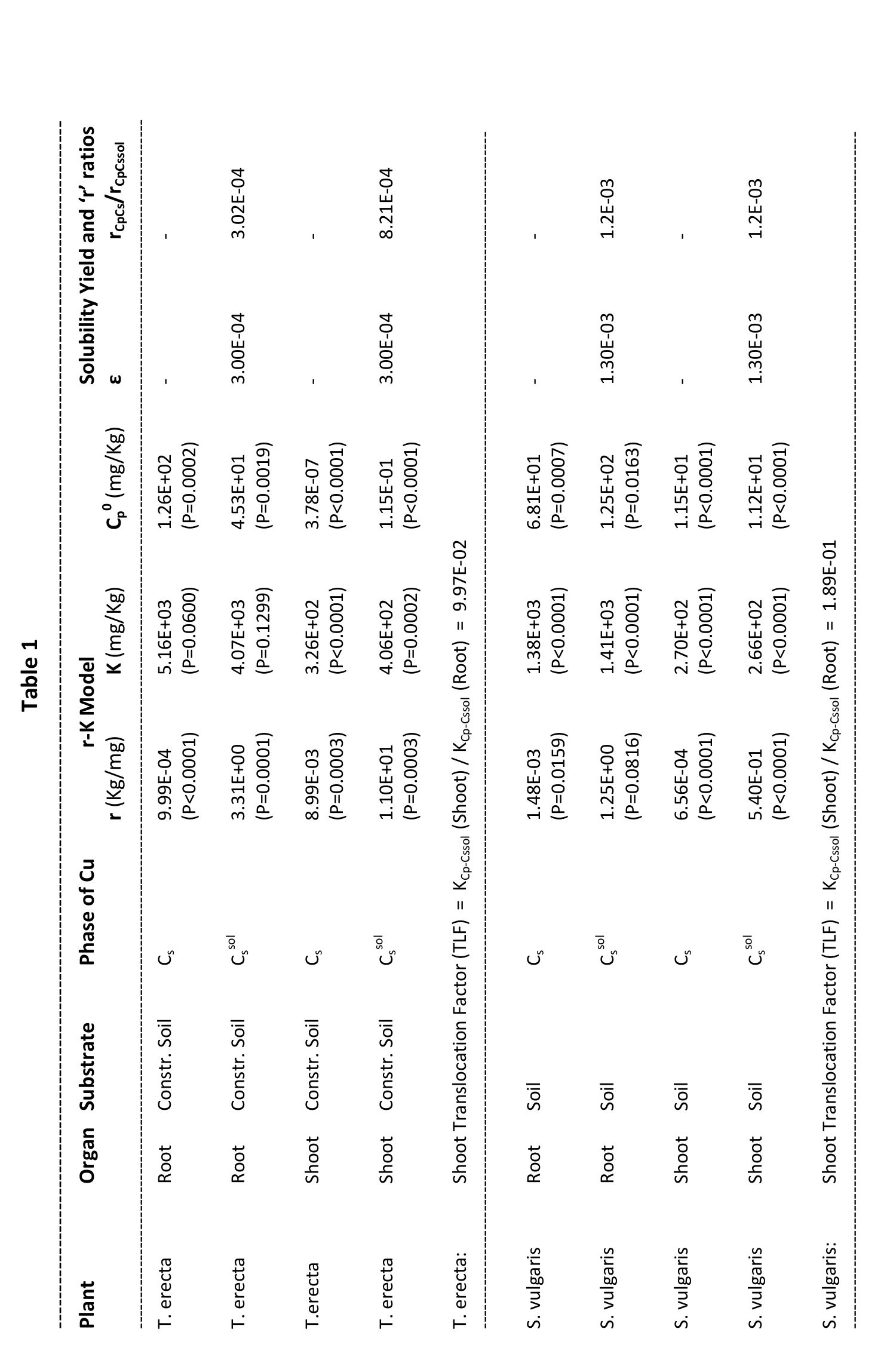}
\end{center}
\label{table1}
\end{figure}

\begin{figure}[h]
\begin{center}
\includegraphics[width=1.15\textwidth]{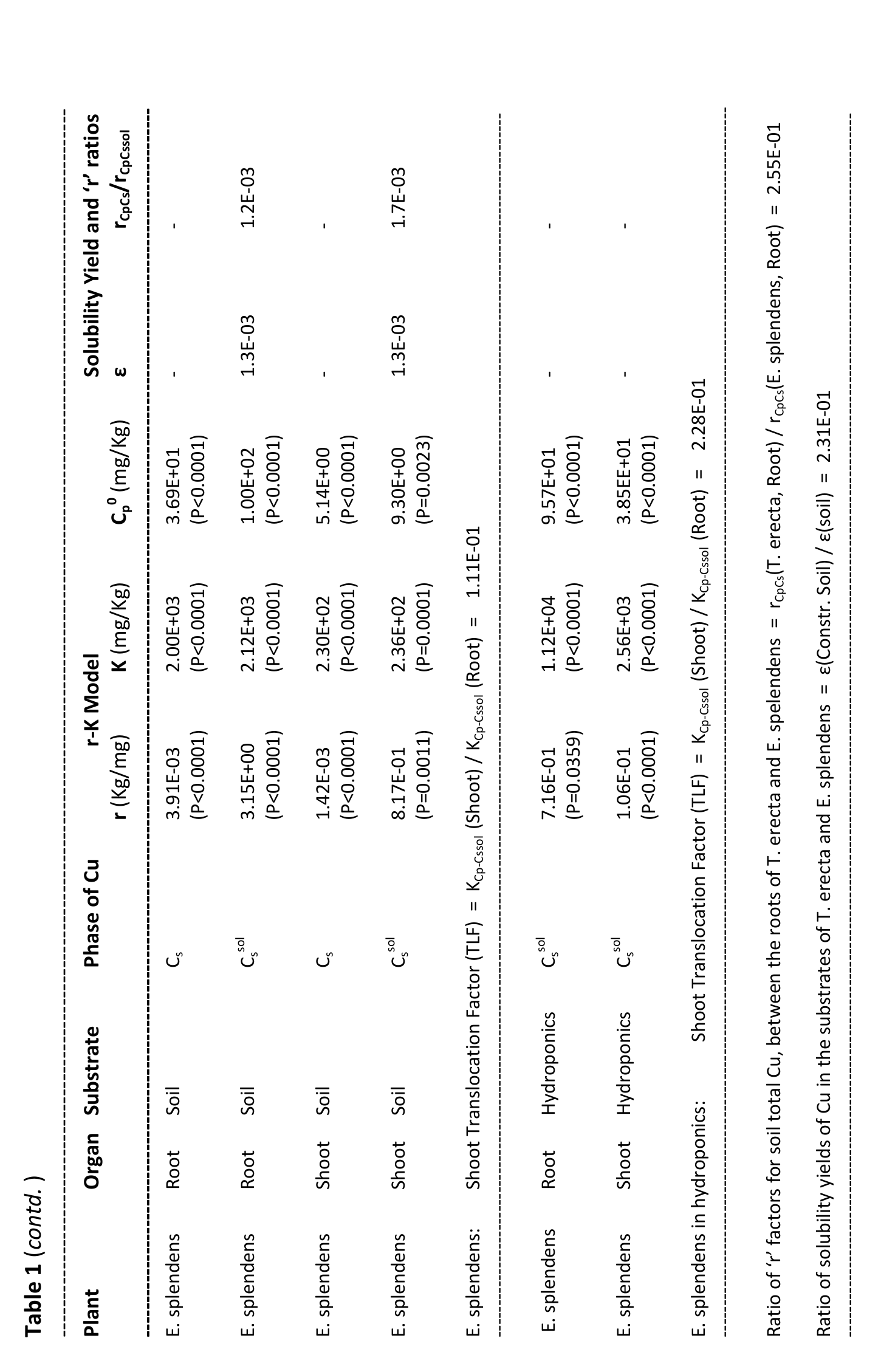}
\end{center}
\label{table1cont}
\end{figure}

\end{document}